# On the non-standard Lagrangian equations

Aparna Saha[1] and Benoy Talukdar[2]

[1] Department of Physics, Visva-Bharati University, Santiniketan 731235, India

[2] Nutan Palli, Bolpur 731204, India

Email. aparna_phyvb@yahoo.co.in and binoy123@bsnl.in

**Abstract.** In an attempt to look for the root of nonstandard Lagrangians in the theories of the inverse variational problem we introduce a logarithmic Lagrangian (LL) in addition to the so-called reciprocal Lagrangian (RL) that exists in the literature. The equation of motion that follows from the RL could easily be solved by using its first or Jacobi integral. This is not, however, true for the similar equation resulting from the LL. We make use of a method of factorization to find the particular solutions for equations following from both RL an LL and subsequently derive a novel approach to obtain their general solutions from the particular ones. The case studies presented by us include the modified Emden-type equation for which we also construct a new expression for the Lagrangian function (NL). We point out that NL is not related to RL by a total derivative or gauge term. In spite of that NL leads to the same equation of motion as that given by the RL.

## 1. Introduction

For conservative systems, there is an elegant formulation of classical mechanics known as the Lagrangian formulation [1]. The Lagrangian function, $L$, for a system is defined to be the difference between the kinetic and potential energies expressed as a function of generalized coordinates, $q_i(t)$, and velocities, $\dot{q}_i(t)$. Here the overdot denotes differentiation with respect to time $t$. For a system of $N$ particles $q_i = \{q_1, q_2, ... q_N\}$ and $\dot{q}_i = \{\dot{q}_1, \dot{q}_2, ... \dot{q}_N\}$.

The Lagrangian function satisfies the so-called Euler-Lagrange equation given by

$$\frac{d}{dt}\left(\frac{\partial L}{\partial \dot{q}_i}\right) - \frac{\partial L}{\partial q_i} = 0 \ . \qquad (1)$$

The equations that result from application of (1) to a particular Lagrangian are the so-called equations of motion. This is the direct problem of the calculus of variation which provides a framework for the derivation of (1). As opposed to this, one can define an inverse problem [2] that begins with the equation of motion and then constructs a Lagrangian function consistent with the variational principle. While the direct problem is the conventional one, we need to make use of highly sophisticated mathematical techniques [3] to deal with the inverse problem.

Fortunately, there exists a solution of the inverse problem which is intermediate in complexity between the direct and conventional inverse problems. This solution is obtained by coupling the Jacobi integral [1] with the method of characteristics [4] as applied to first-order partial differential equations. Here the Lagrangian is constructed using the first integral of the equation of motion rather than the equation of motion. The first integral is also called the constant of the motion. The Lagrangian function $L(x, \dot{x})$ for a one-dimensional system having a constant of the motion $c(x, \dot{x})$ can be found from [5]

$$L(x, \dot{x}) = \dot{x}\int^{\dot{x}} \frac{c(x, \xi)}{\xi^2}d\xi \ . \qquad (2)$$

Many physical information of a system are encoded in the Lagrangian functions [6]. Also these functions in conjunction with Ritz optimization procedure [7] can be used to find approximate analytical solutions of physically important nonlinear evolution equations [8]. These facts have motivated a large number of studies on the direct and inverse problems of Lagrangian mechanics. In the recent past, Cariñena, Rañada and Santander [9] introduced the concept of non-natural Lagrangians which involve neither the kinetic term nor the potential function but lead to equations of motion that represent physically interesting nonlinear systems. All results were presented without explicit reference to the inverse problem of the calculus of variations. Relatively recently, Musielak [10], and Cieśliński and Nikiciuk [11] used a kind of 'Bootstrap Model' to construct expressions for standard (kinetic energy minus potential energy) as well as non-standard (non-natural) Lagrangians (NSLs) with a view to identify classes of equations of motion that admit a Lagrangian representation. In ref.11 the non-standard Lagrangians of ref.9 were termed as reciprocal Lagrangians.

The object of the present work is to look into the root of non-standard Lagrangians in the theories of inverse variational problem by making use of (2) and also to provide an uncomplicated method to solve the equations of motion that follow from them. In section 2 we make use of (2) to introduce the so-called reciprocal Lagrangian. Interestingly, the proposed solution of the inverse problem introduces in a rather natural way a logarithmic type of NSLs. We find that both reciprocal and logarithmic Lagrangians via Euler-Lagrange equations lead to Liénard-type nonlinear differential equations which differ in their constant coefficients. By calculating Jacobi integrals [1] we show that equations resulting from the reciprocal Lagrangian are easily solvable. This is, however, not true for equations following from the logarithmic Lagrangian. In section 3 we envisage some case studies and obtain a few equations of physical interest. In this context we also present a new result for the NSL of the modifird Emden-type equations [12]. In section 4 we provide solutions of the equations by using an uncomplicated method. Finally, in section 5 we make some concluding remarks with a view to summarize our outlook of the present work.

## 2. Non-standard Lagrangians

For the equation $\ddot{x}(t) = 0$ representing the motion of a free particle, both $c_1 = 1/\dot{x}(t)$ and $c_2 = \dot{x}(t)$ can be regarded as constants of the motion. From the solution of the inverse variational problem in (2), the Lagrangians corresponding to $c_1$ and $c_2$ can be obtained as $L_1(x,\dot{x}) = 1/\dot{x}$ and $L_2(x,\dot{x}) = \dot{x}\ln\dot{x}$. It may be noted that Lagrangians $\dot{x}\ln\dot{x}$ and $\ln\dot{x}$ via Euler-Lagrange equation lead to the same equation of motion. Thus we shall work with

$$L_1(x,\dot{x}) = 1/\dot{x} \text{ and } L_2(x,\dot{x}) = \ln\dot{x} \quad (3)$$

As a simple deformation of (3) we introduce the Lagrangians

$$L_1^d(x,\dot{x}) = 1/(\dot{x} + f(x)) \quad (4a)$$

and

$$L_2^d(x,\dot{x}) = \ln(\dot{x} + f(x)) \quad (4b)$$

where $f(x)$ is a well-behaved function of $x$. Clearly, (4a) stands for the inverse Lagrangian while (4b) represents the corresponding logarithmic one. Thus we see that the root of the nonstandard Lagrangians lies in (3) which was obtained by using the techniques of inverse variational problem. The form in (4a), first introduced in Ref.9, was applied to the study of nonlinear systems related to the generalized Riccati equations. Therefore, it will be interesting to study the properties of the equations of motion that result from the new Lagrangian in (4b). The results in (4a) and (4b) via the Euler-Lagrange equation lead to the equations of motion

$$\ddot{x} + \frac{3}{2}\dot{x}f' + \frac{1}{2}ff' = 0 \quad (5a)$$

and

$$\ddot{x} + 2\dot{x}f' + ff' = 0 \quad (5b)$$

respectively. The prime on $f$ denotes differentiation with respect to $x$. Both equations represent Liénard-type nonlinear systems. One may try to solve them by using their first integrals. In order that we calculate the Jacobi integrals [1] corresponding to $L_1^d(.)$ and $L_2^d(,)$, and find

$$J_1^d = -\frac{f + 2\dot{x}}{(f + \dot{x})^2} \quad (6a)$$

and

$$J_2^d = \frac{\dot{x}}{f + \dot{x}} - \ln(f + \dot{x}) . \quad (6b)$$

From these equations we see that (6a) as obtained using $L_1^d(.)$ can be inverted to write $\dot{x}$ as a function of $J_1^d$ and $x$ while (6b) involves $\dot{x}$ essentially in non-algebraic way. It is rather curious to note that although the differential equations (5a) and (5b) are of the same type their first integrals differ in analytic properties.

## 3. Case studies : NSLs and equations of motion

We shall consider here the particular forms of the nonstandard Lagrangians (4a) and (4b), and construct some physically important equations of motion from them. For the choice $f = kx^n$ with $k$, a time-independent constant, the reciprocal and logarithmic Lagrangians lead to

$$\ddot{x} + \frac{3}{2}knx^{n-1}\dot{x} + \frac{1}{2}nk^2x^{2n-1} = 0, \quad (7a)$$

$$\ddot{x} + 2knx^{n-1}\dot{x} + nk^2x^{2n-1} = 0 \quad (7b)$$

and

$$J_1^d = -\frac{kx^n + 2\dot{x}}{(kx^n + \dot{x})^2}, \quad (8a)$$

$$J_2^d = \frac{\dot{x}}{kx^n + \dot{x}} - \ln(kx^n + \dot{x}). \quad (8b)$$

Here $n = 1, 2, 3 \ldots$ etc.

For $n = 1$ the equations of motion and Jacobi integrals are given by

$$\ddot{x} + \frac{3}{2}k\dot{x} + \frac{1}{2}k^2x = 0, \quad (9a)$$

$$\ddot{x} + 2k\dot{x} + k^2x = 0 \quad (9b)$$

and

$$J_1^d = -\frac{kx + 2\dot{x}}{(kx + \dot{x})^2}, \quad (10a)$$

$$J_2^d = \frac{\dot{x}}{kx + \dot{x}} - \ln(kx + \dot{x}). \quad (10b)$$

Clearly, both (9a) and (9b) represent equations of motion for the damped harmonic oscillator; the first system is over-damped and second one is critically damped. These equations can be solved by simple quadrature. The solution of (9a) may also be found from the Jacobi integral. But it is not possible to obtain the solution (8b) using (9b). This serves as a typical example that a solvable equation cannot be solved using its first integral.

It is well known that the Lagrangian for the damped harmonic oscillator is explicitly time dependent [13]. In the past there were many attempts to construct the time-independent Lagrangian function for dissipative systems and thus accommodate them in the frame of action principle in the usual manner. For example, in a remarkable work Bateman [14] found a time-independent Lagrangian by doubling the degree of freedom of the system. Relatively recently, the same problem was studied by taking recourse to the use of fractional calculus [15]. It is remarkable that the non-standard Lagrangians from which the dissipative equations (9a) and (9b) followed are explicitly time independent by choice.

For $n = 2$ the equations of motion and Jacobi integrals are given by

$$\ddot{x} + 3kx\dot{x} + k^2x^3 = 0, \quad (11a)$$

$$\ddot{x} + 4kx\dot{x} + 2k^2x^3 = 0 \quad (11b)$$

and

$$J_1^d = -\frac{kx^2 + 2\dot{x}}{(kx^2 + \dot{x})^2}, \quad (12a)$$

$$J_2^d = \frac{\dot{x}}{kx^2 + \dot{x}} - \ln(kx^2 + \dot{x}). \quad (12b)$$

Both (10a) and (10b) stand for the modified Emden-type equations [12] of the form

$$\ddot{x} + \alpha x\dot{x} + \beta x^3 = 0 \quad (13)$$

with $\beta = \frac{\alpha^2}{9}$ for (11a) and $\beta = \frac{\alpha^2}{8}$ for (11b).

Equations (11a) and (11b) appear in a number of applicative contexts. For example, they occur in the study of equilibrium configurations of spherical gas cloud under the mutual attraction of its molecules and subject to the laws of thermodynamics [16] and in the modeling of the fusion of pallets [17].

A set of first integrals for (11a) and (11b), other than the Jacobi integrals in (12a) and (12b), can be obtained by writing each of them as a system of coupled first-order differential equations. For example, (11a) is equivalent to

$$\dot{x} = y, \quad y = y(t) \quad (14a)$$

and

$$\dot{y} = -3kxy - k^2x^3. \quad (14b)$$

Combining (14a) and (14b) we find

$$yy' + 3kxy + k^2x^3 = 0. \quad (15)$$

Here, as before, the prime denotes differentiation with respect to $x$. This equation can be integrated to write the constant of the motion as

$$c(x, \dot{x}) = \frac{\dot{x} + kx^2}{\sqrt{2\dot{x} + kx^2}}. \quad (16)$$

From (2) and (16) we get a new result

$$L = NL = \sqrt{2\dot{x} + kx^2}. \quad (17)$$

The Lagrangian in (17) is not connected to the reciprocal one by a gauge term but both of them give the same equation of motion. Such Lagrangians are often called alternative Lagrangians. Two important physical consequences for the existence of alternative Lagrangians of a physical system are (i) there can arise ambiguities in the association of symmetries and conserved quantities and (ii) different Lagrangians can give rise to widely different Hamiltonian systems such that the same classical system can lead to entirely different quantum mechanical systems [18].

For (11b) we found the constant of the motion as

$$c(x,\dot{x}) = \ln(\dot{x} + kx^2) + \frac{kx^2}{\dot{x} + kx^2}. \quad (18)$$

From (2) and (18) we verified that the Lagrangian corresponding to this first integral is the same as that found from (4b). Thus, as opposed to (11a), (11b) does not admit an alternative Lagrangian representation.

## 4. Solutions of the linear and nonlinear equations

In an interesting work Péréz and Rosu [19] showed that for polynomial nonlinearities particular solutions of Liénard-type differential equations can be found by factorizing the second-order system into two first-order operators. Reyes and Rosu [20] found that it is a nontrivial problem to construct the general solution from the particular one. We shall derive an uncomplicated method to find general solutions of our equations by the use of factorization method.

The generic form of the Liénard-type differential equation is given by

$$\ddot{x} + g(x)\dot{x} + F(x) = 0 \quad (19)$$

where $g(x)$ and $F(x)$ are well defined polynomials in $x$. In the factorization method of finding a particular solution of (19) one first writes it in the form

$$(D - \phi_2(x))(D - \phi_1(x))x = 0, \ D = \frac{d}{dt} \quad (20)$$

such that

$$\ddot{x} - (\phi_1 + \phi_2 + \frac{d\phi_1}{dx}x)\dot{x} + \phi_1\phi_2 = 0. \quad (21)$$

Comparison of (19) and (21) yields

$$g(x) = -(\phi_1 + \phi_2 + \frac{d\phi_1}{dx}x) \quad (22a)$$

and

$$\phi_1\phi_2 = \frac{F(x)}{x}. \quad (22b)$$

The values of $\phi_1$ and $\phi_2$ are now chosen with some pre-factors or multiplicative constants so as to satisfy (22b). Subsequently, the multiplicative constants in the $\phi_i$'s are determined by taking recourse to the use of (22a). The particular solution of (19) is then obtained from the first-order differential equation

$$(D - \phi_1)x = 0. \quad (23)$$

We now describe our method to construct the general solution from a given particular primitive. Before applying it to complicated nonlinear equations we would like to illustrate the procedure by dealing with the exactly solvable equations in (9a) and (9b).

For (9a), equations corresponding to (22a) and (22b) read

$$(\phi_1 + \phi_2 + \frac{d\phi_1}{dx}x) = -\frac{3}{2}k \quad (24)$$

and

$$\phi_1\phi_2 = \frac{1}{2}k^2. \quad (25)$$

Let us assume

$$\phi_1 = \frac{1}{2}ak \text{ and } \phi_2 = a^{-1}k \quad (26)$$

where $a$ is a constant. From (24) and (26) we have $a = -1$ and $-2$. Using the value $a = -1$ in $\phi_1$ we solve (23) to get a particular solution of (9a) in the form

$$x = ce^{-\frac{kt}{2}} \quad (27)$$

where $c$ is a constant of integration. In the spirit of the method of variation of parameters we now allow the constant $c$ in (27) to become a function of time $t$ [21] such that

$$x = c(t)e^{-\frac{kt}{2}}. \quad (28)$$

This value of $x$ when substituted in (9a) leads to the differential equation

$$2\ddot{c} + k\dot{c} = 0 \quad (29)$$

for $c(t)$. The solution of (29) can now be substituted in (28) to get

$$x = Ae^{-\frac{kt}{2}} + Be^{-kt} \quad (30)$$

where $A$ and $B$ are the constants of integration arising from the solution of (29). A similar consideration when applied to (9b) gives the solution

$$x = (A + Bt)e^{-kt} . \qquad (31)$$

Note that using $a = -2$ we would get the same results as those in (30) and (31).

The first two nonlinear equations in the hierarchy of equations provided by (7a) and (7b) are the modified Emden-type equations in (11a) and (11b). We shall now apply the procedure outlined above to obtain their solutions. For (11a) we have

$$\phi_1 + \phi_2 + \frac{d\phi_1}{dx} = -3kx \qquad (32a)$$

and

$$\phi_1 \phi_2 = k^2 x^2 . \qquad (32b)$$

In view of (32b) we choose

$$\phi_1 = akx \text{ and } \phi_2 = a^{-1}kx \qquad (33)$$

with $a$ an arbitrary constant. Equations (32a) and (33) can be combined to get

$a = -1$ and $-\frac{1}{2}$. For $a = -1$, (23) gives the particular solution

$$x = \frac{1}{kt - c} , \qquad (34)$$

where $c$ is a time-independent constant. We now make $c$ time dependent and write

$$x = \frac{1}{kt - c(t)} . \qquad (35)$$

Substituting (35) in (11a) we obtain a differential equation for $c(t)$ as

$$(kt - c)\ddot{c} + 2\dot{c}^2 - k\dot{c} = 0 . \qquad (36)$$

Albeit nonlinear, (36) can be solved analytically to obtain

$$c(t) = \frac{e^{-4\alpha_1}(kt^2 - k^2\alpha_2^2 - e^{=2\alpha_1})}{2(t - \alpha_2)} \qquad (37)$$

where $\alpha_1$ and $\alpha_2$ are new constants of integration. If we take $\alpha_1 = \alpha_2 = 0$ we get

$$c(t) = \frac{kt^2 - 1}{2t} \qquad (38)$$

which when used in (34) leads to well-known solution

$$x = \frac{2t}{1 + kt^2} . \qquad (39)$$

The modified Emden-type equation with $\beta = \frac{\alpha^2}{9}$ $\beta$ possesses eight parameter Lie point symmetries [22]. This might be the reason why our method worked so smoothly to obtain the solution. This is, however, not the case with (11b) where $\beta = \frac{\alpha^2}{8}$. Thus a useful check on the fidelity of our method will consist in applying it to (11b).

It is straightforward to verify that the particular solution of (11b) obtained by the factorization method is the same as that found in (34) for (11a). To get the general solution we thus substitute (35) in (11b) to write

$$(c - kt)\ddot{c} - 2\dot{c}^2 + 8k\dot{c} - 8k^2 = 0 . \qquad (40)$$

Unlike (36), (40) cannot be integrated analytically. We thus try to find a solution of it by introducing a change of variable written as

$$c(t) = g(t) + kt . \qquad (41)$$

From (40) and (41) we get

$$g\ddot{g} - 2\dot{g}^2 + 4k\dot{g} - 2k^2 = 0 \qquad (42)$$

which is again of the same form as that in (40). However, if we choose to work with $\dot{g} = 0$, (42) can be solved to get

$$g(t) = e^{\frac{1}{4}\left(-\frac{c_1}{k^2} - 4\,erf^{-1}\left(\frac{2i\sqrt{e^{\frac{c_1}{2k^2}}}(t+c_2)}{\sqrt{\pi}}\right)\right)} \qquad (43)$$

where $c_1$ and $c_2$ are arbitrary constants of integration, and $erf^{-1}$ stands the inverse error function. Since $\dot{g} = 0$ implies that $\dot{c} = k$, the reciprocal of the result in (43) refers only to a solution of (11b) which is not the most general one.

In the recent past Chandrasekar et.al [23] solved (10b) by treating it as a Hamiltonian system and then taking recourse to the use of a canonical transformation. Interestingly, our solution of (10b) for $c_1 = 1$ and $c_2 = -it_0$ is in exact agreement with that of (16) in ref. 21 where it was also not taken as a general solution. The general solution was found by substituting the result in the equation for the canonical transformation of $x(t)$. We, however, follow a different viewpoint to find the general solution.

From (35) and (41) we see that $\dfrac{1}{g(t)}$ represents the solution of (11b) only when the first derivative of the variational parameter is fixed at $k$. The question is how can one relax this condition and find the most general solution of the equation.

It is easy to verify that the transformation

$$x(t) = \frac{1}{k}\frac{d}{dt}\ln y(t) \qquad (44)$$

reduces (11a) in the linear form

$$\ddot{y} = 0 \qquad (45)$$

and the solution of (45) when substituted in (44) gives the exact solution of (11a). Unfortunately, no such transformation can be found to reduce (11b) in the linear form. In this situation we suggest that the choice $y(t) = \dfrac{1}{g(t)}$ (46) together with (45) will at least lead to an approximate general solution of (11b). Following this viewpoint we obtain

$$x(t) = 2\,ierf^{-1}(z)$$
$$\exp\left[\frac{1}{2}\left(\frac{1}{2k^2} + 2\{erf^{-1}(z)\}^2\right)\right] \qquad (47)$$

where

$$z = \frac{2ikt}{\sqrt{\pi}}\exp\left(\frac{1}{2k^2}\right). \qquad (48)$$

The solution (47) of (11b) is in agreement with that found in ref.23 where the authors claimed their result as an exact one. Thus the analysis presented in this section firmly establishes that the factorization technique supplemented by the method of variation of parameters as proposed by us provides a powerful approach to solve important Newtonian equations that follow from both RL and LL type nonstandard Lagrangians.

## 5. Concluding remarks

All fundamental equations of modern physics can be derived from the corresponding Lagrangian functions. This direct problem of the calculus of variations is well documented in many textbooks and monographs [24]. On the other hand. many physicists are hardly familiar with the inverse problem of Lagrangian mechanics [2]. As we have already noted, here one begins with the equations of motion and then constructs their Lagrangians by a strict mathematical procedure [25]. The general solution of the inverse vaiational was given by Helmholtz [26]. Darboux [27] proved that in the case of one-dimensional problems the Lagrangian always exists. But the Lagrangian description is highly non-unique and provides an opportunity to choose Lagrangian functions in ad hoc fashions. The nonstandard Lagrangian of the reciprocal type [9] provides a typical example in respect of this.

In this work we envisaged a study to find the root of the RL in the theory of inverse variational problem. This provided us an opportunity to introduce another non-standard Lagrangian, namely, the logarithmic Lagrangian which we abbreviated as LL. In writing the result for the LL we have chosen to work with the deformation of $c_2 = \ln \dot{x}$ rather than $c_2 = \dot{x}\ln \dot{x}$. Because of the non-uniqueness of the Lagrangians for one-dimensional systems, this particular choice is not a limitation of our approach. Also we noted that Lagrangians obtained from the deformation of $c_2 = \dot{x}\ln \dot{x}$ lead to physically uninteresting equations of motion. However, we found that the complete description of physical systems require the knowledge of both RL and LL. We demonstrated this with particular emphasis on the equations of the damped harmonic oscillator and Modified Emden-type equation. The latter represents a nonlinear ordinary differential equation for which it is difficult to devise a standard method of solution. We have demonstrated that these equations can be satisfactorily solved by taking recourse to the use of a method of factorization in conjunction with a variant of the Lagrange's method of variation of parameters. A very interesting feature of the factorization method is that the set of equations found from the use of RL and LL has exactly the same particular solution. But their general solutions are quite different. We conclude by noting that Lagrangian formalism developed by us is based a definitive solution of the inverse variational problem of classical dynamics. Our method of solving the nonlinear equations which result the derived Lagrangians is also quite straightforward for application to other equations of nonlinear dynamics.